\documentstyle[dina4,epsf,capt2dd]{article}
\def\ppmm#1#2{\pm{\textstyle{#1\vrule height 1.5ex width 0pt
    \atop #2 \vrule depth 0.4ex width 0pt}}}
\begin{document}
{\pagestyle{empty}
\setlength{\oddsidemargin}{0pt}
\setlength{\evensidemargin}{0pt}
\setlength{\marginparwidth}{2.4truecm}
\setlength{\marginparsep}{0pt}
\setlength{\topmargin}{0pt}
\setlength{\headheight}{0pt}
\setlength{\headsep}{0pt}
\setlength{\topskip}{0pt}
\setlength{\textwidth}{16.4truecm}
\footheight 12pt \footskip 24pt
\renewcommand{\baselinestretch}{1.0}
\parskip = 4pt plus 0.10fill
\let\thepage\relax
\def\centerformat{%
    \pretolerance = 2000 \tolerance = 3000 \hbadness 500
    \parindent = 0pt \parskip = 0pt \catcode"0D = 5
    \rightskip = 20pt plus 1fill\leftskip  = 20pt plus 1fill
    \spaceskip = 0.35em \xspaceskip = 0.45em \parfillskip = 0pt
    \hyphenpenalty = 1000 \exhyphenpenalty = 1000}
\def\:{.\kern 2pt\relax}
\font\sixrm=cmr6
\font\egtrm=cmr8
\def\title#1{{\centerformat{%
    \Large \bf
    #1\par}}%
    \vskip 2pt\vskip\parskip}
\def\authors#1{{\centerformat{\rm #1\par}}}
\def\inst#1{{\centerformat{\it #1\par}}\vskip \parskip}
\newcount\fnnumber
\def\fn#1#2{\/%
    \expandafter\ifx\csname FnN#1\endcsname\relax%
    \global\advance\fnnumber 1%
    \expandafter\xdef\csname FnN#1\endcsname%
    {\the\fnnumber}%
    \insert\footins{\interlinepenalty=\interfootnotelinepenalty
    \leftskip=0pt\rightskip=0pt\parfillskip=0pt plus 1fill
    \spaceskip=2.5pt plus 1.5pt minus 1pt
    \xspaceskip=3pt plus 2pt minus 1pt
    \baselineskip = 0.8\normalbaselineskip \lineskip = \normallineskip
    \pretolerance=2000\tolerance=3000
    \hyphenpenalty=500\exhyphenpenalty=50
    \hangindent 4.1mm\hangafter 1
    \noindent\hbox to 4.1mm{\hss$^{\sixrm\the\fnnumber}$\kern 0.67mm}%
    \vrule width 0pt height 1\baselineskip\egtrm #2}%
    \fi$^{\csname FnN#1\endcsname}\!$}
\vbox to 0pt{\vss\noindent\hfill DESY 97-194\break\hbox{}\par}

\title{Determination of the Michel Parameters $\rho$, $\xi$, and $\delta$\\
in $\tau$-Lepton Decays with $\tau\rightarrow\rho\nu$ Tags}
\vspace*{2mm}
\inst {The ARGUS Collaboration}

\authors {%
 H\:Albrecht,
 T\:Hamacher,
 R\:P\:Hofmann,
 T\:Kirchhoff,
 R\:Mankel\fn{Kola}{Now at Institut f\"ur Physik,
                 Humboldt-Universit\"at zu Berlin, Germany.},
 A\:Nau,
 S\:Nowak\fn{IfH}{DESY, IfH Zeuthen, Germany.},
 D\:Re\ss ing,
 H\:Schr\"oder,
 H\:D\:Schulz,
 M\:Walter\fn{IfH}{},
 R\:Wurth
}
\inst {DESY, Hamburg, Germany}

\authors {%
 C\:Hast,
 H\:Kapitza,
 H\:Kolanoski\fn{Kola}{},
 A\:Kosche,
 A\:Lange,
 A\:Lindner,
 M\:Schieber,
 T\:Siegmund,
 H\:Thurn,
 D\:T\"opfer,
 D\:Wegener}
\inst {Institut f\"ur Physik\fn{DO}{%
 Supported by the German
 Bundesministerium f\"ur Forschung und Technologie, under contract
 number 054DO51P.},
 Universit\"at Dortmund,
 Germany}

\authors {%
 C\:Frankl,
 J\:Graf,
 M\:Schmidtler\fn{Sci}{Now at Caltech, Pasadena, USA.},
 M\:Schramm,
 K\:R\:Schubert,\\
 R\:Schwierz,
 B\:Spaan,
R\:Waldi}
\inst{Institut f\"ur Kern- und Teilchenphysik\fn{DR}{%
Supported by the German Bundesministerium f\"ur Forschung
und Technologie, under contract number 056DD11P.},
Technische Universit\"at Dresden, Germany}

\authors {%
 K\:Reim,
 H\:Wegener}
\inst{Physikalisches Institut\fn{ER}{%
 Supported by the
 German Bundesministerium f\"ur Forschung und Technologie, under
 contract number 054ER12P.}, Universit\"at Erlangen-N\"urnberg,
 Germany}

\authors {%
 R\:Eckmann,
 H\:Kuipers,
 O\:Mai,
 R\:Mundt,
 T\:Oest,
 R\:Reiner,
 A\:Rohde,
 W\:Schmidt-Parzefall}
\inst{II. Institut f\"ur Experimentalphysik, Universit\"at Hamburg,
 Germany}

\authors {%
 J\:Stiewe,
 S\:Werner}
\inst{Institut f\"ur Hochenergiephysik\fn{HD}{%
 Supported by the
 German Bundesministerium f\"ur Forschung und Technologie, under
 contract number 055HD21P.}, Universit\"at Heidelberg, Germany}

\authors {%
 K\:Ehret,
 W\:Hofmann,
 A\:H\"upper,
 K\:T\:Kn\"opfle,
 J\:Spengler}
\inst{Max-Planck-Institut f\"ur Kernphysik, Heidelberg, Germany}

\authors {%
 P\:Krieger%
\fn{To}{University of Toronto, Toronto, Ontario, Canada.},
 D\:B\:MacFarlane%
\fn{Mg}{McGill University, Montreal, Quebec, Canada.},
 J\:D\:Prentice\fn{To}{},
 P\:R\:B\:Saull\fn{Mg}{},
 K\:Tzamariudaki\fn{Mg}{},
 R\:G\:Van~de~Water\fn{To}{},
 T.-S\:Yoon\fn{To}{}}
\inst {Institute of Particle
 Physics\fn{CDN}{%
 Supported by the Natural Sciences and Engineering
 Research Council, Canada.}, Canada}

\authors {%
 M\:Schneider,
 S\:Weseler}
\inst{Institut f\"ur Experimentelle Kernphysik\fn{KA}{%
 Supported by the
 German Bundesministerium f\"ur Forschung und Technologie, under
 contract number 055KA11P.}, Universit\"at Karlsruhe, Germany}

\authors {%
 M\:Bra\v cko,
 G\:Kernel,
 P\:Kri\v zan,
 E\:Kri\v zni\v c,
 G\:Medin\fn{lj}{On leave from University of Montenegro, Yugoslavia},
 T\:Podobnik,
 T\:\v Zivko
}
\inst{Institut J. Stefan and Oddelek za fiziko\fn{YU}{Supported
 by the Ministry of Science and Technology of the Republic of
 Slovenia and the Internationales B\"uro KfA,
 J\"ulich.}, Univerza v Ljubljani, Ljubljana, Slovenia}

\authors {%
 V\:Balagura,
 S\:Barsuk,
 I\:Belyaev,
 R\:Chistov,
 M\:Danilov,
 V\:Eiges,
 L\:Gershtein,
 Yu\:Gershtein,
 A\:Golutvin,
 O\:Igonkina,
 I\:Korolko,
 G\:Kostina,
 D\:Litvintsev,
 P\:Pakhlov,
 S\:Semenov,
 A\:Snizhko,
 I\:Tichomirov,
 Yu\:Zaitsev}
\inst{Institute of Theoretical and Experimental Physics\fn{ITEP}{%
 Partially supported by Grant MSB300 from the International
Science Foundation.},
Moscow, Russia}

\vskip 0pt plus 1fill
\eject
}%

\parindent0.0pt
\begin{center}
{\bf Abstract}
\end{center}
{\small
Using the ARGUS detector at the $e^{+}e^{-}$ storage ring DORIS II, we
have measured the Michel parameters $\rho$, $\xi$, and $\xi\delta$ for
$\tau^{\pm}\rightarrow l^{\pm} \nu\bar\nu$ decays in $\tau$-pair
events produced at center of mass energies in the region of the
$\Upsilon$ resonances.
Using $\tau^\mp \to \rho^\mp \nu$ as spin analyzing tags, 
we find
$\rho_{e}=0.68\pm 0.04 \pm 0.08$, 
$\xi_{e}= 1.12 \pm 0.20 \pm 0.09$,
$\xi\delta_{e}= 0.57 \pm 0.14 \pm 0.07$, 
$\rho_{\mu}= 0.69 \pm 0.06 \pm 0.08$, 
$\xi_{\mu}= 1.25 \pm 0.27 \pm 0.14$ and
$\xi\delta_{\mu}= 0.72 \pm 0.18 \pm 0.10$. 
In addition, we report the combined ARGUS results on
$\rho$, $\xi$, and $\xi\delta$ using this work und 
previous measurements.}

\section{Introduction}
A general ansatz to describe purely leptonic $\tau$ decays
is the well-known formalism by Michel \cite{04}. This formalism
includes extensions of the Standard Model such as scalar and tensor
coupling to left- or righthanded leptons. The derived differential
width in the $\tau$ rest frame neglecting radiative corrections 
and terms proportional to $m^{2}_{l}/m^{2}_{\tau}$ is 
\begin{eqnarray}
\frac{d^{2} \Gamma (\tau^{\pm}\rightarrow l^{\pm}\nu\bar\nu)}{d\Omega\,dx}
   &=& \frac{G^{2}_{F}m^{5}_{\tau}}{192 \pi^{4}} x^{2} \cdot \left[
       3 (1-x)+\frac{2}{3}\rho(4x-3) +6\eta\frac{m_{l}}{m_{\tau}}
       \frac{1-x}{x} \right. \nonumber\\ \label{EqnMic}
   & & \left.\pm P_{\tau} \cos\vartheta\left(\xi (1-x)+\frac{2}{3}
       \xi\delta(4x-3)\right)\right],
\end{eqnarray}

where $x=2 E_{l}/m_{\tau}$ is the scaled energy of the charged lepton,
$P_{\tau}$ the $\tau$ polarization and $\vartheta$ the angle between the 
polarization vector and the charged lepton momentum.  Equation~(\ref{EqnMic})
is valid for vanishing neutrino masses. The Standard Model with a
pure $V-A$ interaction predicts
$\rho = 3/4$, $\eta = 0$, $\xi = 1$, and $\xi\delta = 3/4$.

Up to now, measurements of the Michel parameters $\xi$ and $\xi\delta$ 
have been made by ARGUS, ALEPH, L3, SLD and CLEO \cite{12,03,papers}. In this paper, we present measurements
of $\rho$, $\xi$, and $\xi\delta$ with a new tagging method 
\cite{warsaw} separately
for electrons and muons. The results are obtained using 
tau pairs produced in $e^+ e^- \to \tau^+\tau^- \to
(\rho^\pm\nu_{\tau}) (l^\mp\nu_l\nu_\tau)$. The decay
$\tau^\pm \to \rho^\pm\nu_{\tau}$ is used as a tag, i.e.\ for the event
selection
and as $\tau$ spin analyzer. They
are statistically independent of previous ARGUS measurements
using lepton tags \cite{12} and $a_{1}$-meson tags in its decay into three 
charged pions \cite{03}. 

\section{Method of the Measurement}
\begin{figure}[htb]
\unitlength=1mm
\linethickness{0.4pt}
\begin{picture}(150.00,75.00)
\includegraphics{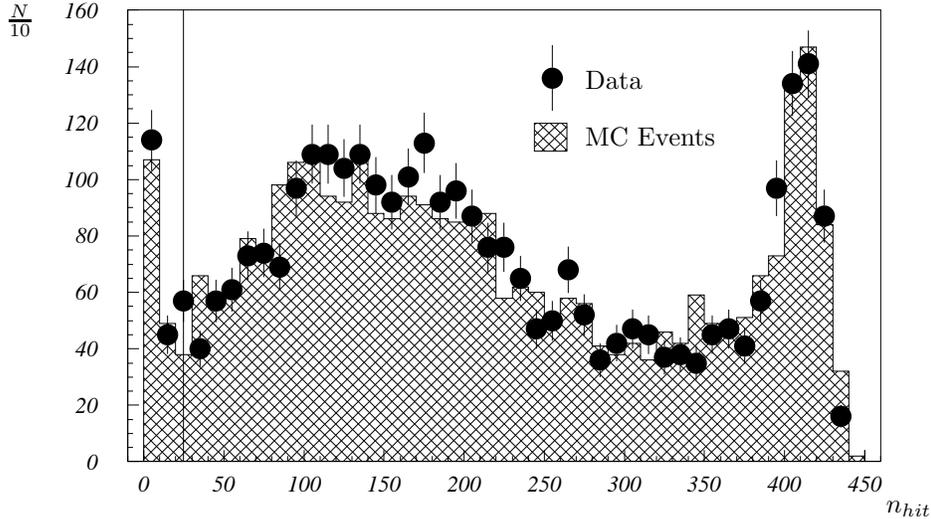}
\put(85.0,59.0){\makebox(0,0)[cl]{Data}}
\put(85.0,51.5){\makebox(0,0)[cl]{MC Events}}
\put(125.0,2.0){\makebox(0,0)[cl]{$n_{hit}$}}
\put(8.0,67.0){\makebox(0,0)[cl]{$N\over{10}$}}
\end{picture}
\caption{Number of matching $(e^{\pm}\nu\bar\nu)(\pi^{\mp}\pi^{0}\nu)$ 
         configurations per event. The points with error bars 
         represent the data. The hatched histogram shows the expectation 
         from the {\sc KoralB/Tauola} Monte Carlo for the decay $(e^{\pm}
         \nu\bar\nu)(\pi^{\mp}\pi^{0}\nu)$ and other contributing 
         $\tau$ channels. The solid line marks $n_{hit} = 25$, the minimum
         number for selected events. Events with a hard initial state photon
         lead mostly to $n_{hit}<25$.\label{FigMat}}
\end{figure}

As can be seen from Eq.~(\ref{EqnMic}),
the measurement of $\xi$ and $\xi\delta$ requires the knowledge 
of the $\tau$ polarization.
At DORIS energies, the average polarization of $\tau$ leptons is
zero, hence no information on $\xi$ and $\xi\delta$ can be
extracted from the analysis of single $\tau$-decays. Only $\rho$ 
and $\eta$ are accessible from the analysis of the momentum
spectrum. However, $\tau$-leptons are produced in pairs resulting
in a spin-spin-correlation. At $\sqrt{s} =$ 10~GeV, $\tau$ pairs are
produced via a virtual photon and both $\tau$ spins are predominantly 
parallel due to the vector coupling, with only a small probability
of $(1-\beta_\tau)/2 \approx m_\tau^2/s \approx 3\%$
for being antiparallel. In the 
analysis described here \cite{07}, we use the decay 
$\tau^{\pm}\rightarrow\rho^{\pm}\nu_{\tau}$ in order to obtain 
information on the $\tau$-spin orientation. The basic
method is similar to 
the one used in the ARGUS analysis with 
$\tau^{\pm}\rightarrow a_{1}^{\pm}\nu_{\tau}$ as spin 
analyzer \cite{03}, and is described in more detail there.\\[2mm]
The performance of various analysis methods has been studied using a Monte 
Carlo simulation.
Tau pair events including radiative photons were produced 
by the {\sc KoralB/Tauola} generator~\cite{08}. These events passed the 
ARGUS detector simulation {\sc Simarg}~\cite{14} and the reconstuction 
program {\sc Arg13}~\cite{02}, and were finally submitted to
the same selection procedure as the data, which is described in 
section~\ref{SecSel} below.\\[2mm] 
The parallel orientation of the two $\tau$ spins is used to relate the
spin of the tau decaying into $l^{\pm}\nu\bar\nu$ to the spin
of the tau decaying into $\rho^{\pm}\nu_{\tau}$.
The spin in the latter decay is related to the $\rho^{\pm}$ direction
because of parity violation with
a neutrino helicity $h_{\nu_{\tau}} = -1$ as supported by
measurements of the sign in $\tau^{\pm}\rightarrow a_{1}^{\pm}\nu_{\tau}$ 
decays \cite{03} and the modulus in 
$\tau^{+}\tau^{-}\rightarrow(\rho^{+}\bar\nu_{\tau})(\rho^{-}\nu_{\tau})$ 
events \cite{15}.
The $\rho$-meson can be 
produced in two helicity states, $h=0$ and
$h= +1$. The state with $h=0$ dominates by a factor of $(1+\beta^*)/(1-\beta^*)
= m_{\tau}^{2}/m_{\rho}^{2}
\approx 5.3$, where $\beta^*$ is the velocity of the $\tau$ in
the $\rho$ rest frame.  Therefore, 
its $\tau$-spin $\rho$-momentum correlation also dominates,
resulting in correlations between the momenta $p_{l}$ of the lepton and 
$p_{\rho}$ of the $\rho$-meson of the same tau pair event.
In a Monte Carlo sample with the Standard Model value $\xi=1$
on the lepton side and $h_{\nu_{\tau}}=-1$ on the $\rho$ side, we obtain a
correlation coefficent $r = {\rm Cov}(p_{l},p_{\rho})/(\sigma(p_{l})\cdot
\sigma(p_{\rho})) = (-6.6\pm 0.4)$\% whereas parity conservation
($h_{\nu_{\tau}}\cdot\xi=0$) leads to $r = (-2.2\pm 0.4)$\% which is
non-zero because of radiation in the initial $e^{+}e^{-}$ state.
A measurement of $r$ is, therefore, a
measurement of $h_{\nu_{\tau}}\cdot\xi$.\\[2mm]
Additional information on $\xi$ and $\xi\delta$
is obtained by (i) using the momentum-momentum 
correlations of the vectors instead of just their absolute values, and
(ii) including 
information on the $\rho$-meson helicity, since the suppressed
$h= +1$ state results in the opposite correlation compared to
the $h=0$ state.  A discrimination between both states is possible 
via the decay angle for 
$\rho^{\pm}\rightarrow \pi^{\pm}\pi^{0}$, since the distributions
for both helicities are different.\\[2mm]
Maximal information on $\rho$,
$\xi$, and $\xi\delta$ is obtained by a single entry maximum likelihood
fit using the selected $(l^{\pm}\nu\bar\nu)(\pi^{\mp}\pi^{0}\nu)$ events
with all their 10 measured quantities, the $e^+e^-$ cms energy $\sqrt{s}$, 
and the momentum vectors of the three observed particles,
$\vec p(l^{\mp})$, 
$\vec p(\pi^{\pm})$, and $\vec p(\pi^{0})$.
The matrix element for the differential cross section 
$e^{+}e^{-}\rightarrow \tau^{+}\tau^{-}\rightarrow
(l^{\pm}\nu\bar\nu) (\pi^{\mp}\pi^{0}\nu)$ 
can be written as
\begin{equation}\label{EqnMat}
\vert {\cal{M}} \vert^{2} = H\, A \, [L_{1} +\rho L_{2} +\eta L_{3}] +
                 h_{\nu_{\tau}} H'_{\alpha} C^{\alpha\beta} 
                 [\xi L'_{1\beta} + \xi\delta L'_{2\beta}] ,
\end{equation}
where the first part is the spin-averaged contribution and the second part 
describes the $\tau^{+}\tau^{-}$-spin correlations. 
$H$ and $H'$ describe the decays $\tau^{\pm}\rightarrow\rho^{\pm}\nu$ and 
$\rho^{\pm}\rightarrow \pi^{\pm}\pi^{0}$.
$L_{1}$, $L_{2}$, $L_{3}$, $L'_{1}$, and $L'_{2}$ describe 
$\tau^{\pm}\rightarrow l^{\pm}\nu\bar\nu$ decays in the most 
general way with zero
neutrino masses. $A$ is the spin-averaged matrix element for $\tau$-pair
production and $C$ the spin-spin correlation matrix ($\alpha,\beta = 1
\ldots 4$). $\vert {\cal M}\vert^{2}$ includes radiative corrections
which are treated in a factorizable way.\\[2mm]
The likelihood function to be maximized contains three variables 
${\theta} = (\rho, \xi, \xi\delta)$ and $N\cdot 10$ measured quantities
$m_{i} = (s_{i},\vec  p_i(l^{\mp}),\vec p_i(\pi^{\pm}),\vec p_i(\pi^{0}))$,
where $i=1\ldots N$, and $N$ is the number of
selected events,
\begin{equation}
L ({\theta}) = \prod\limits_{i=1}^{N} L_{i} ({\theta}\vert m_{i})
                 = \prod\limits_{i=1}^{N} f (m_i\vert{\theta}).
\end{equation}
Since data were taken at different cms energies,
the probability density $f$ is taken as a function of scaled
momenta $p_i/\sqrt s$.  It is
the projection of the full
density $f (m,u\vert {\theta})$ where $u$ denotes
the unobserved
kinematical quantities which include neutrino momenta as well as
photon momenta of initial and final state radiation. 
Furthermore, the
observables $m_{i}$ are obtained as a convolution of the true values $t_{i}$ 
of the same kinematical quantities and the detector resolution
function $\cal{R}$. The final density is given by
\begin{equation}
f(m\vert{\theta}) = \eta(m)
 \cdot{\cal{N}}({\theta})
 \int \vert {\cal{M}}(t,u\vert{\theta})\vert^{2} 
 \cdot{\cal{P}} (t,u) 
 \cdot{\cal{R}}(m\vert t) \,d t\,d u ,
\end{equation}
where
$\cal{P}$ is the phase space distribution,
$\eta(m)$ is the detector acceptance,
and  $\cal{N}({\theta})$ is the normalization factor given
by
\begin{equation}
{\cal{N}}({\theta})\cdot \int \eta(m)
 \cdot\vert {\cal{M}}(t,u\vert{\theta})\vert^{2} 
 \cdot{\cal{P}} (t,u) 
 \cdot{\cal{R}}(m\vert t) \,d t\,d u\,d m = 1 .
\end{equation}
Radiative corrections including initial state radiation and internal
bremsstrahlung in $\tau \to e \nu\bar\nu$ are included in $ {\cal{M}}(t,u\vert{\theta})$.
External bremsstrahlung of the electron is included in the resolution function 
${\cal{R}}(m\vert t)$. 
For the application as a likelihood function, $f(m\vert{\theta})$ is 
rewritten in a more appropriate form. The factorizing radiative corrections
allow to write $\vert{\cal M}(t,u\vert\theta)\vert^{2} \cdot
{\cal P}(t,u) = \sum_{j=0}^{3} \theta_{j} \,F_{j}(t,u)$ with
$\theta_{0}\equiv 1$.
The likelihood of a single event~$i$ is then
\begin{equation} \label{EqnLik}
L_{i}({\theta}\vert m_{i})
    = \frac{\sum\limits_{j=0}^{3}\theta_j \int F_j(t,  u) \, 
      {\cal{R}}(m_i\vert t)\, dt\, du}{
      \sum\limits_{j=0}^{3}\theta_j\int\eta(m) \,F_j(t, u) \,
      {\cal{R}}(m\vert t)\, dt\,du\,dm} ,
\end{equation}
where the acceptance factor has been omitted in the numerator since it
depends only on the observed quantities $m$. The integration over the
neutrino momenta, which are part of the unobserved quantities $u$, is
performed to a large extent analytically, leaving only an uncertainty
in the relative angular orientation of the two tau leptons.
The integration over $t$ and the remaining seven unobserved variables $u$ 
is done by a Monte Carlo method in which for each event 450 kinematical 
configurations are tested for compatibility with the observed set of
observables $m_{i}$. 
The seven $u$ variables in the Monte Carlo integration are
the momentum vector of the initial state photon,
the $\tau$ orientation angle, and the momentum vector of the photon
in $\tau \to e \nu\bar\nu\gamma_2$.
A configuration is ``successful'' if the set of $(m_i,t,u)$ is compatible
with the kinematics of
$e^+ e^- \to \gamma_1 (\pi^\pm \pi^0 \nu) (l^\mp \nu\bar\nu \gamma_2)$
events, i.e.\ if the undetermined neutrino momenta can have physical values. 
\\[2mm]
These 450 tries lead to $n_{hit}$ successful matches
for each event, and the distribution of $n_{hit}$ is shown in
figure~\ref{FigMat}.  The high value of $\langle n_{hit}\rangle$ shows
the effectiveness of the numerical integration chosen, and $n_{hit}$
is large enough for most events to compute $L_i(\theta|m_i)$ with
sufficient precision.  Events with $n_{hit} < 25$ are rejected in the data
selection.
The good agreement between $N(n_{hit})$ for
selected data and for accepted Monte Carlo events supports the validity
of the assumptions used.
\\[2mm]
The integral in the 
denominator of Eq.~(\ref{EqnLik}), for each set of parameters~${\theta}$, 
is also
obtained with the help of a Monte Carlo integration which determines the four
separate integrals used in the sum.\\[2mm]
The validity of the method has been extensively tested by using Monte Carlo
events as pseudo data for determining the parameter $\rho$, $\xi$, and 
$\xi\delta$. Even using pseudo data samples 10 times larger than the 
real data samples, we did not detect any significant biases in the
determined parameter values.

\section{Data Selection\label{SecSel}}
The measurements presented here were performed with the ARGUS
detector at the $e^{+}e^{-}$-storage ring DORIS II. The detector
and its trigger requirements are described elsewhere \cite{02}.
The data sample used was collected between 1983 and 1989 in the
center of mass energy region of the $\Upsilon$ resonances. 
The integrated luminosity used in this analysis is $289$ pb$^{-1}$
corresponding to 265\,000 $\tau$ pairs produced.

Event selection starts with requiring two oppositely charged tracks 
forming a vertex in the interaction region and an opening angle larger 
than 80$^{\circ}$ at the vertex point. Each track must have a 
transverse momentum above 150~MeV/c and point into the barrel region.
We demand one particle to be identified as a lepton. The other track
has to be accompagnied by one or two neutral clusters in its hemisphere
with a minimum angle of 10$^{\circ}$ to the track.
A neutral cluster is defined as energy deposition above 100~MeV in the 
calorimeter. No additional cluster is allowed, neither in the lepton
hemisphere nor in the 10$^{\circ}$ cone around the $\pi^{\pm}$ candidate.
In the single cluster case, accounting for photons from the 
$\pi^{0}$ decay  merging into one cluster, we demand a minimum 
energy deposition of 1~GeV. In the two cluster case, both photons 
are combined to form a $\pi^{0}$. The two photon system 
must have an invariant mass within 100~MeV of the nominal 
$\pi^{0}$ mass and $\chi^{2} \leq$ 9 for the
$\pi^{0}$ mass hypothesis.  The lepton identification was done by a 
likelihood method using specific energy loss, time of flight, amount 
and lateral spread of energy deposition in the calorimeter and hits in 
the muon chambers. More details on this likelihood method are found in 
refs.~\cite{10,11}. To ensure a good lepton identification, a momentum 
above 0.8~GeV/c for electrons and above 1.1~GeV/c for muons as well as 
an electron or muon likelihood higher than 0.8 is required.
The charged pion candidate is required to have a pion likelihood 
above 0.6 using specific energy loss in the drift chamber and 
time-of-flight \cite{02}.

\begin{figure}[htb]
\unitlength=1mm
\linethickness{0.4pt}
\begin{picture}(140.00,80.00)
\includegraphics{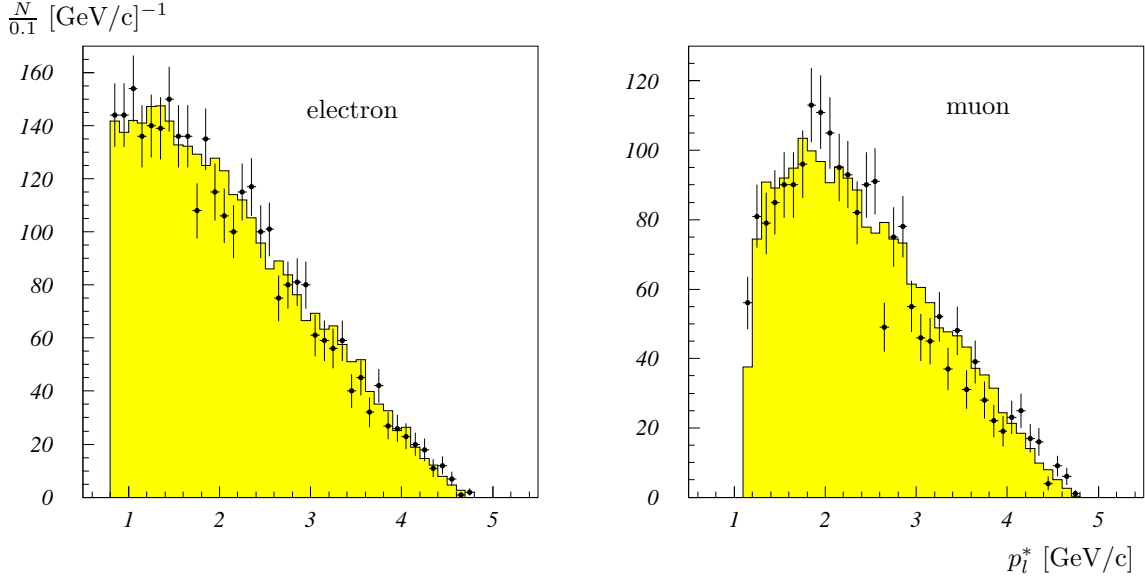}
\put(40.00,60.00){\makebox(0,0)[lc]{electron}} 
\put(125.00,60.00){\makebox(0,0)[lc]{muon}}
\put(150.00,0.00){\makebox(0,0)[rc]{$p^{\ast}_{l}$  [GeV/c]}} 
\put(0.00,72.00){\makebox(0,0)[lc]{$N \over{0.1}$ [GeV/c]$^{-1}$}} 
\end{picture}
\caption{Spectra of scaled momentum $p^{\ast}_{l}= p_{l} \cdot 10~{\rm GeV}/
         \sqrt{s}$ for leptons. Data are shown as dots with error bars
         and accepted Monte Carlo events, generated with $\rho=$ 3/4, as 
         grey shaded area.\label{FigSpe}}
\end{figure}

\begin{figure}[htb]
\unitlength=1mm
\linethickness{0.4pt}
\begin{picture}(150.00,75.00)
\includegraphics{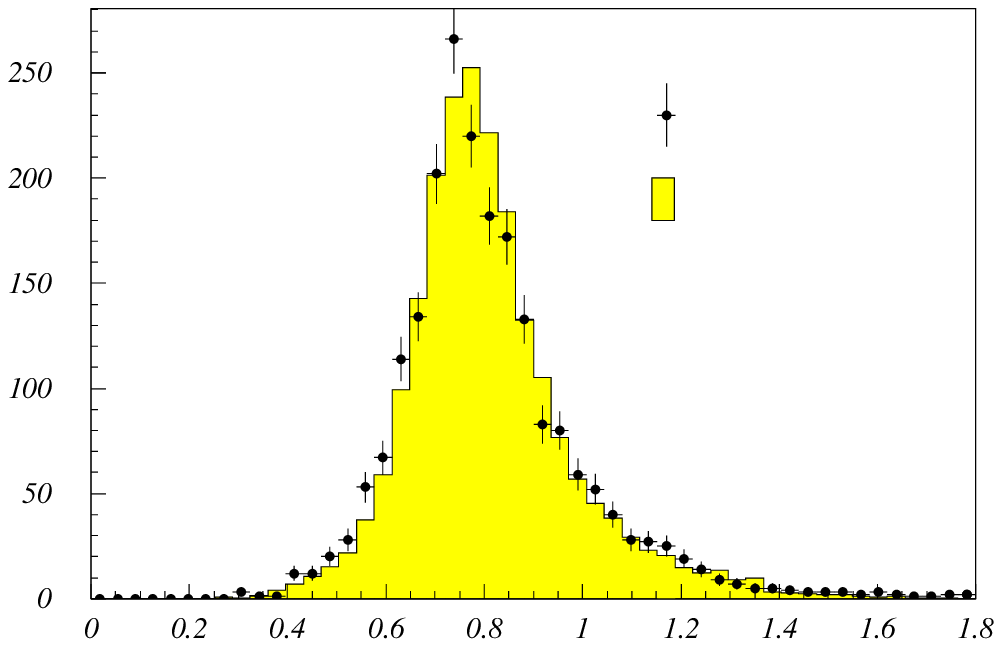}
\put(154.0,0.0){\makebox(0,0)[rc]{$m_{\pi^{\pm}\pi^{0}}$ $\rm [GeV/c^{2}]$}}
\put(20.0,68.0){\makebox(0,0)[cc]{$N\over 0.036$ [GeV/c$^{2}$]$^{-1}$}} 
\put(93.0,53.0){\makebox(0,0)[lc]{Data}} 
\put(93.00,44.5){\makebox(0,0)[lc]{MC Events}} 
\end{picture}
\caption{Invariant mass distribution $m_{\pi^{\pm}\pi^{0}}$ for the 
         $(\pi^{\pm}\pi^{0}\bar\nu)(\mu^{\mp}\nu\bar\nu)$ signature.
         The Monte Carlo spectrum includes background from other
         tau decays. 
         \label{FigMas}}
\end{figure}

To reject background, we apply the following three additional event cuts.
The invariant mass $m_{\pi^{\pm}\pi^{0}}$ has to be smaller than
the $\tau$ lepton mass. This cut suppresses $q\bar q$-events.
The angle between the lepton momentum $\vec{p}_{l}$ and the
$\rho$-meson momentum $\vec{p}_{\rho}$ is not larger than 176$^{\circ}$,
which rejects radiative Bhabha and $\mu$ pair events.
The relation $f_{c}\cdot (m^{2}_{miss}/s - f_{o})^{4} < p_{t}/\sqrt{s}$ 
with $f_{c} = 5$ and $f_{o} = 0.45$ for the single cluster case, and
$f_{c} = 3$ and $f_{o} = 0.4$ for the two cluster case has to be
fulfilled. In here, the missing mass $m_{miss}$ uses only 
charged tracks and the vector sum of transverse momenta $p_{t}$ uses
charged and neutral particles. This cut removes $\gamma\gamma$ and
radiative Bhabha events.\\[2mm]
After these cuts, a data sample of 3176 candidates for $(e^{\pm}\nu\bar\nu)
(\pi^{\mp}\pi^{0}\nu)$ and 2099 candidates for $(\mu^{\pm}\nu\bar\nu)
(\pi^{\mp}\pi^{0}\nu)$ remains. As seen in figure~\ref{FigSpe}, a good 
agreement for lepton momenta is found comparing data and Monte Carlo 
events. However, the data selection criteria do not allow to remove all 
background sources. The selected sample still contains 10\% ($\pi^{\pm}\pi^{0}
\pi^{0}\nu$)($l^{\mp} \nu\bar\nu$) decays, 3.5\% ($K^{\pm}\pi^{0}\nu$)($l^{\mp}
\nu\bar\nu$) decays, $\approx$ 1\% misidentified leptons and around 
2\% background from non-$\tau$ channels, as determined by Monte Carlo. 
Figure~\ref{FigMas} shows the $\pi^{\pm}\pi^{0}$ invariant mass distribution
with good agreement between data and Monte Carlo events. Tau decay fractions 
for the background estimation are taken from ref.~\cite{05}.

\section{Data Analysis}
A first information on $\xi$ is obtained from the $p_{l}$-$p_{\rho}$
correlation coefficient. The selected data sample has $r = -(6.1 \pm 1.2)$\%
leading to $\xi = 0.89 \pm 0.27$ without background corrections.
The finally used likelihood function in Eq.~(\ref{EqnLik}) can conveniently 
include background from other semihadronic $\tau$ decays in the data sample
and from lepton misidentification. The modified likelihood function is
\begin{equation} \label{EqnMod}
L^{\ast}_{i} = (1-\lambda_{a_{1}}-\lambda_{K^{\ast}}) 
\left( (1-\lambda_{f})\cdot L_{i} 
+ \lambda_{f} \cdot L_{i}^{fake} \right) 
+ \lambda_{a_{1}}\cdot L_{i,(\pi^{\pm}\pi^{0}\pi^{0})}
+ \lambda_{K^{\ast}} \cdot L_{i,(K^{\pm}\pi^{0})} ,
\end{equation}
where $\lambda_{a_{1}}$ is the fraction of the 
($\pi^{\pm}\pi^{0}\pi^{0}$)-tags,
$\lambda_{K^{\ast}}$ is the fraction of the ($K^{\pm}\pi^{0}$)-tags and
$\lambda_{f}$ considers pions misidentified as leptons.
Table~\ref{Tab01} gives the used values for $\lambda_{f}$, $\lambda_{
a_{1}}$, and $\lambda_{K^{\ast}}$ as determined by Monte Carlo simulation. 
The additional likelihood functions $L_{i,k}$ for 
$k=(\pi^{\pm}\pi^{0}\pi^{0}\nu)(l^{\mp}\nu\bar\nu)$ and  
$(K^{\pm}\pi^{0}\nu)(l^{\mp}\nu\bar\nu)$
are determined by the same technique as $L_{i}$ for $(\pi^{\pm}\pi^{0}\nu)
(l^{\mp}\nu\bar\nu)$. $L_{i}^{fake}$ is the contribution to the 
likelihood function for misidentified leptons. Non-$\tau$ background has 
not been considered in the likelihood function, because its contribution 
is small and is included in the systematic error 
studies, see Table~\ref{Tab02}.

\begin{table}[htb]
\begin{center}
\begin{tabular}{|l|c|c|}
\hline
  & & \\[-2mm]
  & $(\pi^{\pm}\pi^{0}\nu)(e^{\mp}\nu\bar\nu)$ & $(\pi^{\pm}
                  \pi^{0}\nu)(\mu^{\mp}\nu\bar\nu)$ \\[1mm] 
\hline\hline
  & & \\[-2mm]
$\lambda_{f}$        & $( 0.8\pm0.2)\%$ & $( 1.6\pm0.2) \%$ \\[1mm]
$\lambda_{a_{1}}$    & $(10.1\ppmm{2.8}{0.6})\%$ & $(10.5\ppmm{2.9}{0.6})\%$\\[1mm]
$\lambda_{K^{\ast}}$ & $( 3.7\pm1.0) \%$ & $( 3.4\pm1.0) \%$\\[1mm]
\hline
\end{tabular}
\end{center}
\caption{Monte Carlo determined background levels as 
         used in the likelihood function Eq.~(4).  The asymmetric errors
         are due to the systematic deviation of ARGUS branching fractions from 
         the world average used to calculate these factors.\label{Tab01}} 
\end{table}

The maximum likelihood fits lead to results and statistical errors for
the Michel parameters as shown in Table~\ref{Tab03} below.
Figures \ref{FigMat} and \ref{FigLih} illustrate the quality of the fit and 
the validity of the assumptions made for the determination of the 
kinematical quantities used to calculate the matrix elements. 

\begin{figure}[htb]
\unitlength=1mm
\linethickness{0.4pt}
\begin{picture}(120.00,75.00)(-18.00,0.00)
\includegraphics{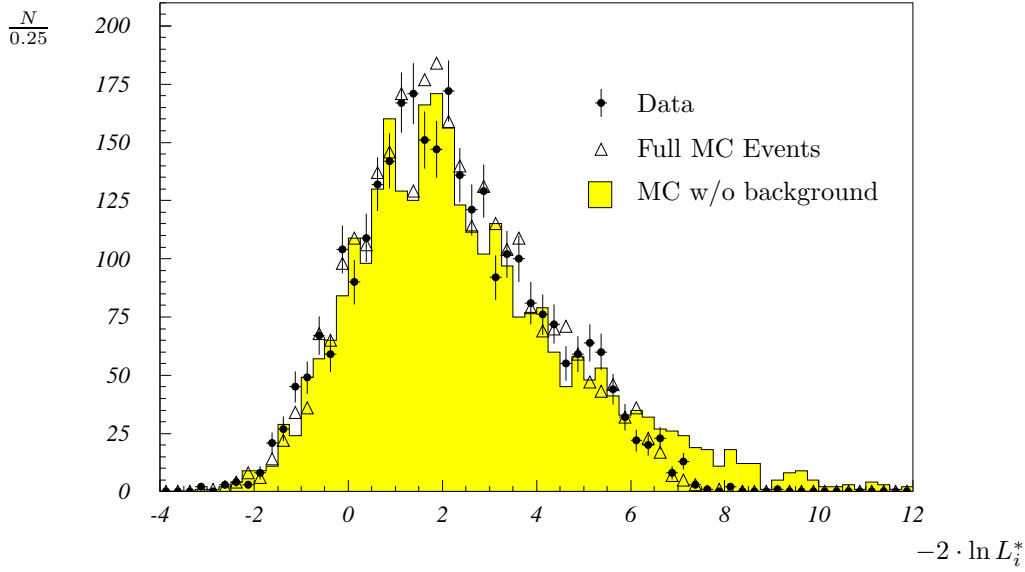}
\put(110.00,0.00){\makebox(0,0)[lc]{$-2\cdot\ln L^{\ast}_{i}$}} 
\put(-5.00,70.00){\makebox(0,0)[rc]{$N \over 0.25$}} 
\put(73.00,60.00){\makebox(0,0)[lc]{Data}} 
\put(73.00,54.00){\makebox(0,0)[lc]{Full MC Events}} 
\put(73.00,48.00){\makebox(0,0)[lc]{MC w/o background}} 
\end{picture}
\caption{The contribution of single events to $-2\ln L_{i}^{\ast}$
         for events with $n_{hit} \geq $ 25.
         The points with error bars represent the data. The open triangles 
         show the expectation from the {\sc KoralB/Tauola} Monte Carlo 
         for the decay $(e^{\pm}\nu\bar\nu)(\pi^{\mp}\pi^{0}\nu)$ and 
         other contributing $\tau$ channels. The shaded area are
         Monte Carlo events without consideration of any background from other $\tau$ 
         decay channels or lepton misidentification.\label{FigLih}}
\end{figure}

Figure~\ref{FigMat} shows the distribution of successful tries to find
a matching initial state to the observed final state comparing data
and Monte Carlo events. Figure~\ref{FigLih} shows the agreement 
in $-2 \ln L^{\ast}_{i}$ for the fit to the data compared 
to the fit to Monte Carlo events generated with Standard Model parameters
described with Eq.~(\ref{EqnMod}). Deviations are found at large values
of $-2\cdot\ln L^{\ast}_{i}$ if we neglect other $\tau$ sources and 
misidentified leptons.

\begin{table}[htb]
\begin{center}
\begin{tabular}{|l|ccc|ccc|}
\hline
 & & & & & & \\[-2mm]
systematic error source
       & \multicolumn{3}{c|}{$(e^{\pm}\nu\bar\nu)(\pi^{\mp}\pi^{0}\nu)$} 
       & \multicolumn{3}{c|}{$(\mu^{\pm}\nu\bar\nu)(\pi^{\mp}\pi^{0}\nu)$}
         \\[1mm]
       & $\Delta (\rho)$ & $\Delta (\xi)$ & $\Delta (\xi\delta)$ 
       & $\Delta (\rho)$ & $\Delta (\xi)$ & $\Delta (\xi\delta)$ \\[1mm] 
\hline
 & & & & & & \\[-2mm]
Monte Carlo statistics &
    $\pm 0.01$ & $\pm 0.05$ & $\pm 0.03$ & 
    $\pm 0.02$ & $\pm 0.07$ &$\pm 0.04$\\[1mm]
Trigger efficiency      &
    $\pm 0.01$  & $\pm 0.01$ &$\pm 0.01$&
    $\approx 0$ & $\pm 0.07$ &$\pm 0.01$\\[1mm]
Branching fractions &
    $^{+0.020}_{-0.003}$&$^{+0.041}_{-0.007}$&$^{+0.024}_{-0.004}$
   &$^{+0.026}_{-0.004}$&$^{+0.044}_{-0.007}$&$^{+0.057}_{-0.009}$\\[1mm]
energy dependence&
    $\pm 0.03$ & $\pm 0.02$ & $\pm 0.02$ & 
    $\pm 0.03$ & $\pm 0.03$ & $\pm 0.02$\\[1mm]
particle identification &
    $^{+0.05}_{-0.06}$ & $\pm 0.03$&$\pm 0.03$ 
   &$\pm 0.04$ & $\pm 0.02$ & $\pm 0.02$\\[1mm]
lepton misidententification &
    $0.005$ &$\pm 0.01$ & $\pm 0.004$ &
    $0.005$ &$\pm 0.02$ & $\pm 0.005$\\[1mm]
background (non $\tau$ physics)&
    $\pm 0.01$ & $\pm 0.03$ & $\pm 0.03$ &
    $\pm 0.01$ & $\pm 0.07$ & $\pm 0.06$\\[1mm]
\hline
 & & & & & & \\[-2mm]
total systematic error &
    $\pm 0.08$  & $\pm 0.09$ & $\pm 0.07$ &
    $\pm 0.08$  & $\pm 0.14$ & $\pm 0.10$\\[1mm]
\hline
\end{tabular}
\end{center}
\caption{\label{Tab02} Contributions to the systematic error}
\end{table}

The systematic errors are summarized in Table~\ref{Tab02} including
limited  Monte Carlo statistics, trigger efficiencies,
branching fractions, particle identification efficiencies, lepton
misidentification,
the influence of background from non-$\tau$ physics
and the center of mass energy dependence of the matrix element
description.
The total error is obtained as a quadratic sum of all errors in 
Table~\ref{Tab02}.

\section{Results and Conclusions}

The Michel parameters as found by the maximum likelihood analysis described above,
and including systematic errors,
are given in Table~\ref{Tab03}. 
\begin{table}[htb]
\begin{center}
\begin{tabular}{|c|c|c|c|}
\hline
 & & & \\[-2mm]
          & $(\pi^{\mp}\pi^{0}\nu)(e^{\pm}\nu\bar\nu)$ &
            $(\pi^{\mp}\pi^{0}\nu)(\mu^{\pm}\nu\bar\nu)$ &
            combined fit \\[1mm]
\hline\hline
 & & & \\[-2mm]
$\rho$      & $0.68 \pm 0.04 \pm 0.07$ & $0.69 \pm 0.06 \pm 0.06$ 
            & $0.68 \pm 0.04 \pm 0.07$\\[1mm]
$\xi$       & $1.11 \pm 0.20 \pm 0.08$ & $1.26 \pm 0.27 \pm 0.14$ 
            & $1.18 \pm 0.16 \pm 0.10$\\[1mm]
$\xi\delta$ & $0.56 \pm 0.14 \pm 0.06$ & $0.73 \pm 0.18 \pm 0.10$ 
            & $0.63 \pm 0.11 \pm 0.08$\\[1mm]
\hline
\end{tabular}
\caption{\label{Tab03} Fitted Michel parameters}
\end{center}
\end{table}

\begin{table}[htb]
\begin{center}
\begin{tabular}{|c|c|c|c|}
\hline
 & & & \\[-2mm]
          & $(\pi^{\mp}\pi^{0}\nu)(e^{\pm}\nu\bar\nu)$ &
            $(\pi^{\mp}\pi^{0}\nu)(\mu^{\pm}\nu\bar\nu)$ &
            combined fit \\[1mm]
\hline\hline
 & & & \\[-2mm]
$\rho$      & $0.72\pm 0.02$ & $0.73\pm 0.02$ 
            & $0.72 \pm 0.02 $\\[1mm]
$\xi$       & $1.12 \pm 0.20\pm 0.09$ & $1.26\pm 0.27\pm 0.14$ 
            & $1.18 \pm 0.16\pm 0.10$\\[1mm]
$\xi\delta$ & $0.59 \pm 0.14\pm 0.07$ & $0.74\pm 0.18\pm 0.10$ 
            & $0.64 \pm 0.11\pm 0.08$\\[1mm]
\hline
\end{tabular}
\caption{\label{Tab04} Fitted results with constrained $\rho$}
\end{center}
\end{table}
The results are in agreement with previous measurements and 
the Standard Model predictions. In addition, the errors are comparable 
to the errors of previous measurements.  The results with $\rho$ 
constrained to the world average of $\rho = 0.742\pm 0.027$ \cite{13} 
are very close to the unconstrained ones; they are given in 
Table~\ref{Tab04} for completeness.
\newpage
All quoted values for $\xi$ and $\xi\delta$ assume a $\tau$ neutrino
helicity $h_{\nu_{\tau}}= -1$.
Using the experimental average $h_{\nu_{\tau}} = -1.011 \pm 0.027$ 
\cite{13} would not change the results.

Assuming lepton universality and combining the results of this analysis
in Table~\ref{Tab03} with previous
ARGUS measurements \cite{09} of $\rho$, $\xi$, and $\xi\delta$, we
obtain
\begin{eqnarray}
\nonumber\rho      &=& 0.731 \pm 0.031\\
\nonumber\xi       &=& 1.03  \pm 0.11 \\ 
\nonumber\xi\delta &=& 0.63  \pm 0.09
\end{eqnarray}

where statistical and systematical errors are combined in quadrature. 
Correlations in the systematic errors of different measurements are
taken into account. The covariance matrix is given by

\begin{center}
\begin{tabular}{cccc}
            & $\rho$  & $\xi$ & $\xi\delta$ \\[1mm]
$\rho$      & $9.54\cdot 10^{-4}$ & & \\[1mm]
$\xi$       & $-2.30\cdot 10^{-5}$ & $1.25\cdot 10^{-2}$ & \\[1mm]
$\xi\delta$ & $-7.88\cdot 10^{-5}$ & $1.03\cdot 10^{-3}$ & $7.84\cdot 10^{-3}$
 \\[1mm]
\end{tabular}
\end{center}

\section*{Acknowledgements}
It is a pleasure to thank U. Djuanda, E. Konrad, E. Michel, and W. Reinsch 
for their competent technical help in running the experiment and processing
the data. We thank Dr. H. Nesemann, B. Sarau, and the DORIS group for
the operation of the storage ring. The visiting groups wish to thank
the DESY directorate for the support and kind hospitality extended to them.


\begin{thebibliography}{99}
\bibitem{04} L. Michel, Proc. Phys. Soc. London, A 63 (1950) 514.

\bibitem{12} ARGUS Collab., H. Albrecht et al., Phys. Lett. B 316 (1993),
             608.
\bibitem{03} ARGUS Collab., H. Albrecht et al., Phys. Lett. B 349 (1995) 576,
             and\\
             M. Schmidtler, ``Bestimmung der Michelparameter $\xi$ und
             $\delta$ in leptonischen $\tau$-Zerf\"allen'',\\
             Dr. rer. nat. Thesis, 
             Universit\"at Karlsruhe, IEKP-KA/94-16 (1994).

\bibitem{papers}ALEPH Collab., Phys. Lett. B 346, 379 (1995);
L3 Collab., Phys. Lett. B 377, 313 (1996);
SLD Collab., Phys. Rev. Lett. 78, 4691 (1997);
CLEO Collab., CLNS-97-1480 (1997).

\bibitem{warsaw}ARGUS Collab.,
contributed paper 
pa~07-098 to the 28th International Conference on High Energy Physics, Warsaw, Poland, July
1996, TUD-IKTP/96-01.

\bibitem{07} M. Schramm, ``Bestimmung der Michelparameter $\xi$ und
             $\delta$ in leptonischen $\tau$-Zerf\"allen mit 
             $\rho$-Mesonen als Spinanalysator'',
             Dr. rer. nat. Thesis, 
             Technische Universit\"at Dresden,\\ TUD-IKTP/97-01 (1997).
\bibitem{08} S. Jadach, J.H. K\"uhn, Z. W\c as, 
             Comp. Phys. Comm. 64 (1991) 275.
\bibitem{14} H. Gennow, ``SIMARG - A Program to simulate the ARGUS 
             Detector'', DESY F15-85-02 (1985).
\bibitem{02} ARGUS Collab., H. Albrecht et al., Nucl. Instr. Meth. A 275
             (1989) 1.
\bibitem{15} ARGUS Collab., H. Albrecht et al., Phys. Lett. {\bf B 337} 
             (1994) 383;\\
             ARGUS Collab., H. Albrecht et al., Z. Phys. {\bf C 58} (1993) 61.
\bibitem{10} S. Weseler, ``Untersuchungen der semileptonischen Zerf\"alle
             von B(5270)-Mesonen mit dem ARGUS-Detektor'', Dr. rer. nat. Thesis, 
             Universit\"at Heidelberg, IHEP-HD/86-02 (1986).
\bibitem{11} B. Fominykh and V. Matveev, ``ARGUS MUON Program'',
             ARGUS software note 14,\\ April 1987. 
\bibitem{05} Particle Data Group, L. Montanet et al., Phys. Rev. D 50,
             1173 (1994) and 1995 off-year update for the 1996 edition. 
\bibitem{13} Particle Data Group, R.M. Barnett et al., Phys. Rev. D 54,
             1 (1996).
\bibitem{09} ARGUS Collab., H. Albrecht et al., ``Physics with ARGUS'',
             Phys. Rep. 276 (1996) 223.
\end{thebibliography}
\end{document}